\documentclass[a4paper,oneside]{article}
\usepackage{graphicx}
\usepackage{multicol}

\def\baselinestretch{0.96}

\def\CPTV{{\begin{picture}(13,0)(0,0)\put(0,0){\tiny CPT}\put(0,0){\line(3,1){13}}\end{picture}}}
\def\CPTVbig{\hbox{\begin{picture}(8,0)(0,0)  \put(0,0){CPT}\put(0,0){\line(3,1){8}}\end{picture}}}

\newcommand{\riga}[1]{\noalign{\hbox{\parbox{\textwidth}{#1}}}\nonumber}
\newcommand{\be}{\begin{equation}}
\newcommand{\ee}{\end{equation}}
\newcommand{\ba}{\begin{array}}
\newcommand{\ea}{\end{array}}

\newcommand{\eV}{\,{\rm eV}}

\def\Red  {}

\def\Black{}
\def\Blue {}
\newcommand{\eq}[1]{~(\ref{eq:#1})}

\newcommand{\NP}{Nucl. Phys.}
\newcommand{\PRL}{Phys. Rev. Lett.}
\newcommand{\PL}{Phys. Lett.}
\newcommand{\PR}{Phys. Rev.}

\newcommand{\fig}[1]{~{\rm \ref{fig:#1}}}

\newcommand{\lascia}[1]{}

\newcommand{\dmatm}{\Delta m^2_{\rm atm}}
\newcommand{\dmbatm}{\Delta\bar{m}_{\rm atm}^2}

\def\circa#1{\,\raise.3ex\hbox{$#1$\kern-.75em\lower1ex\hbox{$\sim$}}\,}
\makeatletter
\def\art{\@ifnextchar[{\eart}{\oart}}
\def\eart[#1]#2#3#4#5#6{{\rm #2}, {\em #3 \bf #4} {\rm (#6) #5} ({\em #1})}
\def\hepart[#1]#2{{\rm #2, \em#1}}
\newcommand{\oart}[5]{{\rm #1}, {\em #2 \bf #3} {\rm (#5) #4}}

\newcounter{alphaequation}[equation]
\def\thealphaequation{\theequation\hbox to
0.6em{\hfil\alph{alphaequation}\hfil}}
\def\eqnsystem#1{
\def\@eqnnum{{\rm (\thealphaequation)}}
\def\@@eqncr{\let\@tempa\relax \ifcase\@eqcnt \def\@tempa{& & &} \or
  \def\@tempa{& &}\or \def\@tempa{&}\fi\@tempa
  \if@eqnsw\@eqnnum\refstepcounter{alphaequation}\fi
\global\@eqnswtrue\global\@eqcnt=0\cr}
\refstepcounter{equation} \let\@currentlabel\theequation \def\@tempb{#1}
\ifx\@tempb\empty\else\label{#1}\fi
\refstepcounter{alphaequation}
\let\@currentlabel\thealphaequation
\global\@eqnswtrue\global\@eqcnt=0 \tabskip\@centering\let\\=\@eqncr
$$\halign to \displaywidth\bgroup \@eqnsel\hskip\@centering
$\displaystyle\tabskip\z@{##}$&\global\@eqcnt\@ne
\hskip2\arraycolsep\hfil${##}$\hfil& \global\@eqcnt\tw@\hskip2\arraycolsep
$\displaystyle\tabskip\z@{##}$\hfil
\tabskip\@centering&\llap{##}\tabskip\z@\cr}
\def\endeqnsystem{\@@eqncr\egroup$$\global\@ignoretrue} \makeatother

\begin{document}
\twocolumn[
\centerline{hep-ph/0201134 \hfill CERN--TH/2002--3\hfill IFUP--TH/2002--3}
\vspace{5mm}
\Black
\vspace{0.5cm}
\centerline{\LARGE\bf\Red Interpreting the LSND anomaly:}
\centerline{\LARGE\bf\Red sterile neutrinos or CPT-violation
or$\ldots$?\footnote{}}
 \medskip\bigskip\Black
   \centerline{\large\bf Alessandro Strumia$^\dagger$}\vspace{0.2cm}
   \centerline{\em Theoretical physics division,
 CERN, CH-1211 Geneva 23, Switzerland }\vspace{0.4cm}
\vspace{5mm}
\Blue\centerline{\large\bf Abstract}
 \begin{quote}\large
We first study how
sterile neutrinos  can fit the $5\sigma$ $\bar{\nu}_\mu\to \bar{\nu}_e$ LSND
anomaly: 2+2 solutions are strongly disfavoured by
solar and atmospheric data,
while 3+1 solutions can still give a poor fit
(for a specific range of oscillation parameters,
to be tested by MiniBooNE).
If  MiniBooNE will see no $\nu_\mu\to \nu_e$ transitions, 
we will have a hint for CPT violation.
Already now, unlike sterile neutrinos,
CPT-violating neutrino masses can accomodate all safe and unsafe data.
We study how much CPT must be conserved according to atmospheric and K2K data and
list which CPT-violating signals could be discovered
by forthcoming solar and long-baseline experiments.
 \Black
 \end{quote}
 \vspace{5mm}]
 \footnotetext[1]{\Red In the addendum at pages \pageref{in}--\pageref{out}
we update our results including the first data  from KamLAND and WMAP, which
disfavour the CPT-violating and `3+1' solutions.
\Black}
\footnotetext[2]{On leave from dipartimento di Fisica
 dell'Universit\`a di Pisa and INFN.}
\Black

\noindent
Oscillations between the three Standard Model (SM) neutrinos are described by
two independent squared neutrino mass differences,
allowing to explain only two of the three neutrino anomalies
(atmospheric~\cite{atmexp}, solar~\cite{sunexp}  and LSND~\cite{LSND})
as oscillations.
A joint fit is not possible even if one trusts only the safest
data from atmospheric,
solar and reactor~\cite{reactorexp}
neutrino experiments:
the the up/down atmospheric asymmetries
and a $\sim 50\% $ disappearance of solar $\nu_e$.
Most global fits of neutrino data drop the LSND anomaly because 
the other ones are considered as more solid.
In quantitative terms, we have a
$8\sigma$ solar anomaly
(although it can be reduced to $5\sigma$ by dropping solar model predictions),
a $14 \sigma$ atmospheric anomaly and a
$5\sigma$ LSND anomaly\footnote{\label{3vs5}
The $\bar{\nu}_\mu\to\bar{\nu}_e$ LSND anomaly is presented as an evidence for a $\mu\to e $
oscillation probability of $(0.264 \pm 0.081)\%$~\cite{LSND},
that differs from zero only by slightly more than $3\sigma$.
However, from a table of the likelihood ${\cal L}$, obtained from the LSND collaboration 
and computed on an event-by-event basis,
we read
$$\Delta\chi^2=\chi^2_{\rm SM}-\chi^2_{\rm best} =  -2 \ln\frac{{\cal L}_{\rm best}}{{\cal L}_{\rm SM}} = 29\qquad\hbox{rather than $\sim 10$.}$$
A reanalysis of LSND data that chooses stronger cuts
obtains $\Delta\chi^2 = 47$
(eq.\ (2.7) of~\cite{LSND2}).
These large $\Delta\chi^2$ mean that the LSND anomaly cannot be due to a statistical fluctuation.
It is not clear which data really contain the LSND evidence.
Apparently, some mark of oscillations that cannot be summarized by the number of $\bar{\nu}_e$ events 
is hidden in the full LSND data, maybe in the energy distribution.
}.
The `number of standard deviations' is here na\"{\i}vely computed as 
$(\Delta\chi^2)^{1/2}=(\chi^2_{\rm SM} - \chi^2_{\rm best})^{1/2}$,
where $\chi^2_{\rm best}$ is the $\chi^2$ value corresponding to the best-fit oscillation,
and $\chi^2_{\rm SM}$ corresponds to massless SM neutrinos.

In section~\ref{sterile} we discuss
how and how well oscillations with 
extra sterile neutrinos
can fit the LSND anomaly~\cite{sterile}.
In particular we study
which one of the two different kind of four-neutrino spectra
(3+1 or 2+2) is favoured by the present data, and by an eventual future
confirmation of the LSND data.
Taking into account the recent SNO result~\cite{sunexp}
an extra sterile neutrino can improve the situation
only in the 3+1 scheme, and even this case
does not allow to fully reconcile all data.

This situation suggests to look for
alternative interpretations of the LSND anomaly.
One possibility is that either the atmospheric or solar or LSND anomaly
is not due to oscillations.
Various mechanisms (even unplausible ones)
can fit the data as well as oscillations~\cite{alternative,KK}.

\begin{figure}[t]
 $$
  \includegraphics[width=50mm,height=4cm]{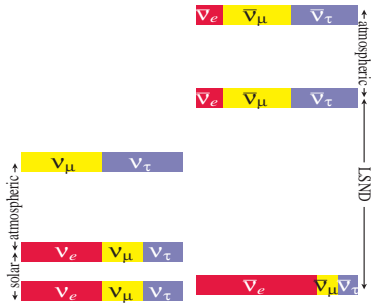} $$
   \caption[]{\em The CPT-violating spectrum proposed in~\cite{MY}.
\label{fig:CPTspettro}}
 \end{figure}

Using only oscillations,
all data
can be consistently fitted by the CPT-violating neutrino
spectrum illustrated in
fig.\fig{CPTspettro}. 
This solution was proposed in~\cite{MY} when
the initial $2.6\sigma$ LSND hint for $\nu_\mu\to \nu_e$~\cite{LSNDinitial} decreased down to $0.6\sigma$,
leaving an anomaly only in $\bar{\nu}_\mu\to \bar{\nu}_e$~\cite{LSND}.
Unlike sterile neutrinos, this solution also satisfies
(unsafe?) bounds from nucleosynthesis and SN1987A~\cite{MY,unsafe?}.
Despite the lack of theoretical grounds,
this speculation is interesting because can be tested soon.
If CPT violation were the right answer,
MiniBooNE~\cite{MiniBooNE} (the experiment designed to test LSND, looking for
${\nu}_\mu\to{\nu}_e$)
will not see the LSND oscillations;
a $\bar\nu_\mu\to \bar\nu_e$ experiment
is needed to directly test this possibility.
If CPT is badly violated as in fig.\fig{CPTspettro}, one generically expects
detectable CPT-violating signals in atmospheric and solar oscillations.
In any case it remains interesting to constrain CPT-violation in neutrino masses.
In section~\ref{CPT}
we compute the present bounds and list the possible CPT-violating
signals and surprises that could appear in
forthcoming solar and long-baseline experiments.

\section{Sterile neutrinos}\label{sterile}
The sterile neutrino can be used to generate either the LSND or the solar or the atmospheric anomaly.

\subsubsection*{3+1 neutrinos}
Within this scheme the sterile neutrino is employed to generate the LSND anomaly.
In fact, in the jargon 3+1 indicates that the additional sterile neutrino
is separated by the large LSND mass gap
from the 3 active neutrinos, separated among them only by the small 
solar and atmospheric mass differences.
A theoretical remark is in order.
If the $4\times 4$ neutrino mass matrix $m_{ii'}$ ($i=\{\ell,s\}$ and $\ell = \{e,\mu,\tau\}$) has the
na\"{\i}ve  form
$$m_{\ell s} = m_{s\ell}=\theta_{\ell s} m_{ss},\quad
m_{s s} = m_{\rm LSND},\quad m_{\ell\ell'}\ll m_{ss}$$
the sterile neutrino induces a contribution to the solar mass splitting of order\footnote{More precisely,
assuming $\theta_{\ell s}\ll 1$, maximal atmospheric mixing and $\theta_{13}=0$, and
taking into account the larger atmospheric mass splitting, one has
$\delta \Delta m^2_{\rm sun} = (\theta_{es}^2 + \theta_{\perp s}^2 )^2\Delta m^2_{\rm LSND}$
where $\theta_{\perp s} \approx (\theta_{\mu s}- \theta_{\tau s})/\sqrt{2}$.}
$$\delta \Delta m^2_{\rm sun} \sim \Delta m^2_{\rm LSND} \sin^2
2\theta_{\rm LSND} \approx 10^{-(3\div 1)}\eV^2$$
that is too large in most of the region allowed by solar and LSND data.
One needs either a cancellation or a mass matrix of the special 
`approximatively rank one'
form
$m_{ii'} \simeq \theta_{is} \theta_{i's} m_{\rm LSND} $.

\medskip

\begin{figure*}[t]
$$
\includegraphics[width=70mm]{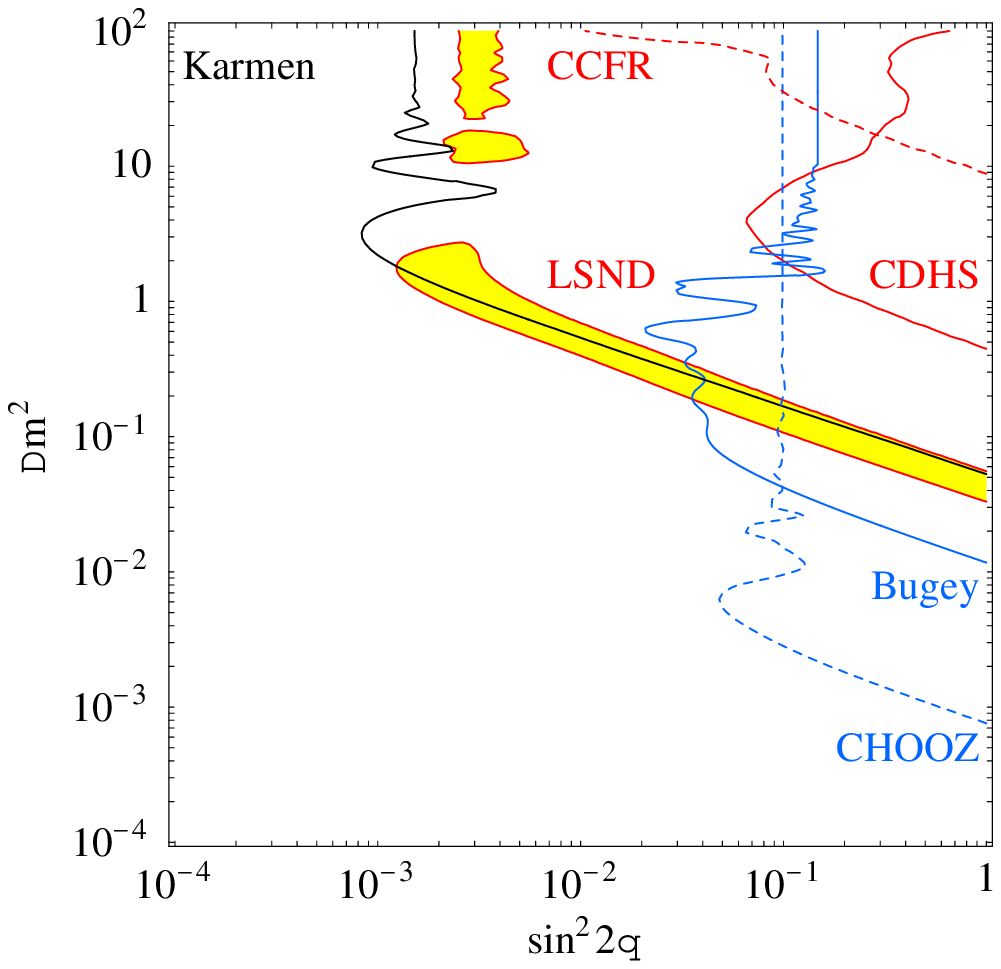} \hspace{1cm}
\includegraphics[width=70mm]{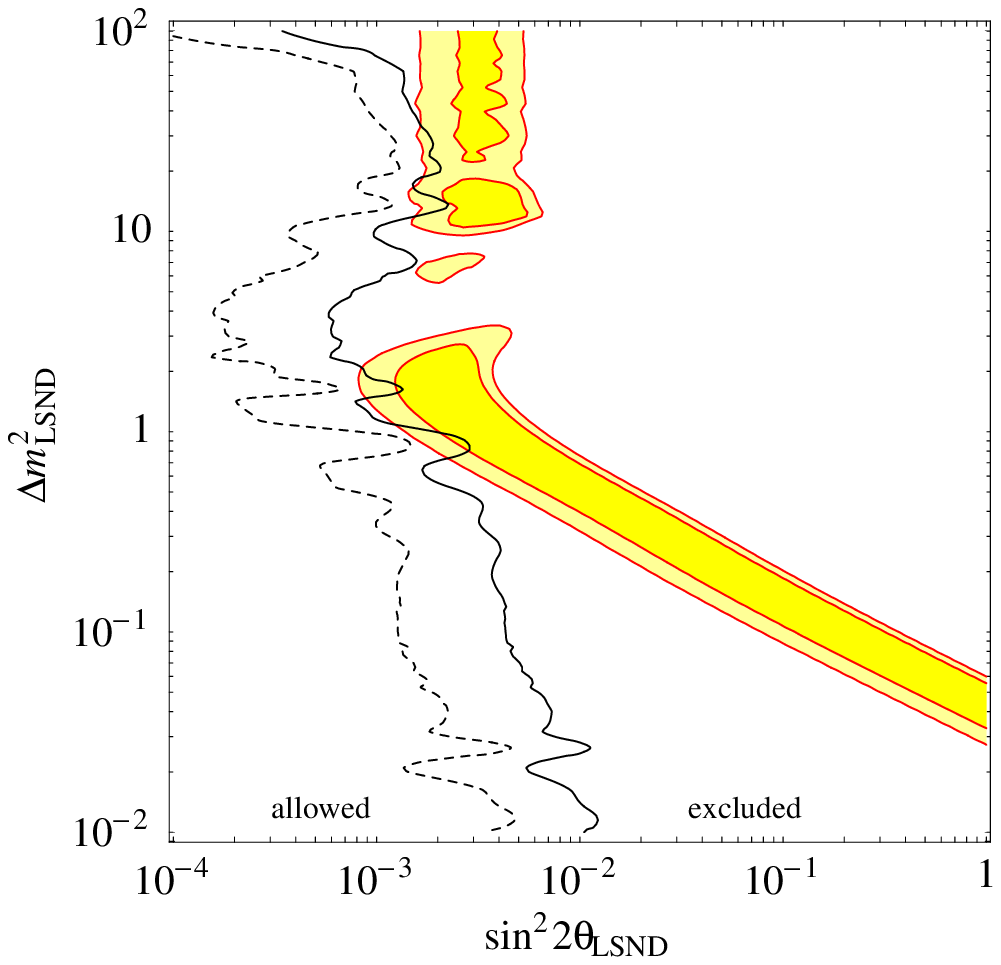}$$
\parbox{\columnwidth}{\caption[]{\em $90\%$ CL regions from Karmen, CDHS, CCFR, {\sc Bugey}, {\sc Chooz} and
LSND (shaded). 
The mixing angle $\theta$ on the horizontal axis 
is  different for the different experiments.
\label{fig:exps}}}\hfill
\parbox{\columnwidth}{\caption[]{\em 
The LSND region at $90\%$ and $99\%$ CL, 
compared with the $90\%$ (dashed line) and $99\%$ CL (continuous line)
combined exclusion bounds from data in fig.\fig{exps} and SK.
\label{fig:split}}}
\end{figure*}

Even ignoring this potential theoretical problem,
3+1 oscillations present a phenomenological problem, because predict that
$\nu_\mu\to \nu_e$ oscillations at the LSND frequency
proceed trough
$\nu_\mu\to \nu_s \to \nu_e$
and $\nu_{e,\mu}\to \nu_s$ are strongly constrained by disappearance experiments.
More precisely, keeping only oscillations at the dominant LSND frequency 
$$S\equiv \sin^2 (\Delta m^2_{\rm LSND} L/4E_\nu)$$ one has
\begin{eqnarray*}
P(\nu_e \to \nu_e) &=& 1-S \sin^2 2\theta_{e s} \\
P(\nu_\mu \to \nu_\mu) &=& 1-S  \sin^2 2\theta_{\mu s} \\
P(\nu_e \to \nu_\mu) &=& S \sin^2 2 \theta_{\rm LSND} 
\end{eqnarray*}
with $\theta_{\rm LSND} \approx \theta_{es}\theta_{\mu s}$, or more precisely~\cite{3+1old}
\begin{equation}\label{eq:qqq}
\sin^2 2 \theta_{\rm LSND}=\frac{1}{4} \sin^2 2\theta_{es}\sin^2 2\theta_{\mu s}.
\end{equation}
The $\theta_{e s}$ mixing angle is constrained by {\sc Bugey}, {\sc Chooz}~\cite{reactorexp}, SuperKamiokande (SK)
atmospheric data~\cite{atmexp}
and the $\theta_{\mu s}$ mixing angle by SK, CDHS and CCFR~\cite{CDHS}.
Furthermore $\nu_\mu\to \nu_e$ oscillations are also directly constrained by Karmen~\cite{Karmen}.
Fig.\fig{exps} illustrates how accurately we reproduce such bounds\footnote{We 
used the SK atmospheric results~\cite{atmexp} after $79$ kton$\cdot$year (55 data), 
K2K~\cite{K2K} (at the moment K2K finds 44 events, versus an expected no-oscillation signal of $64\pm 6$ events),
the latest solar results 
from Homestake, Gallex, SAGE, GNO, SK, SNO (49 data), 
the final {\sc Bugey} (60 data), {\sc Chooz} (14 data), 
CDHS (15 data), CCFR (15 data), {\sc Karmen} and LSND results.
We use the likelihoods computed by the {\sc Karmen} and LSND collaborations
on an event-by-event basis.
We have not included data from {\sc Macro}~\cite{atmexp} (that confirms the atmospheric anomaly)
and from earlier atmospheric experiments
because are less statistically significant than SK.
The data are combined by multiplying all likelihoods ${\cal L}$ 
(i.e.\ by summing all $\chi^2=-2\ln{\cal L}$).
At $\Delta m^2 \circa{>}10\eV^2$ {\sc Chooz} and {\sc Bugey} bounds
could be considered as not fully trustable because 
limited by the theoretical error on the total $\bar{\nu}_e$ fluxes
generated by reactors.
}.

\begin{figure*}[t]
$$\includegraphics[width=70mm]{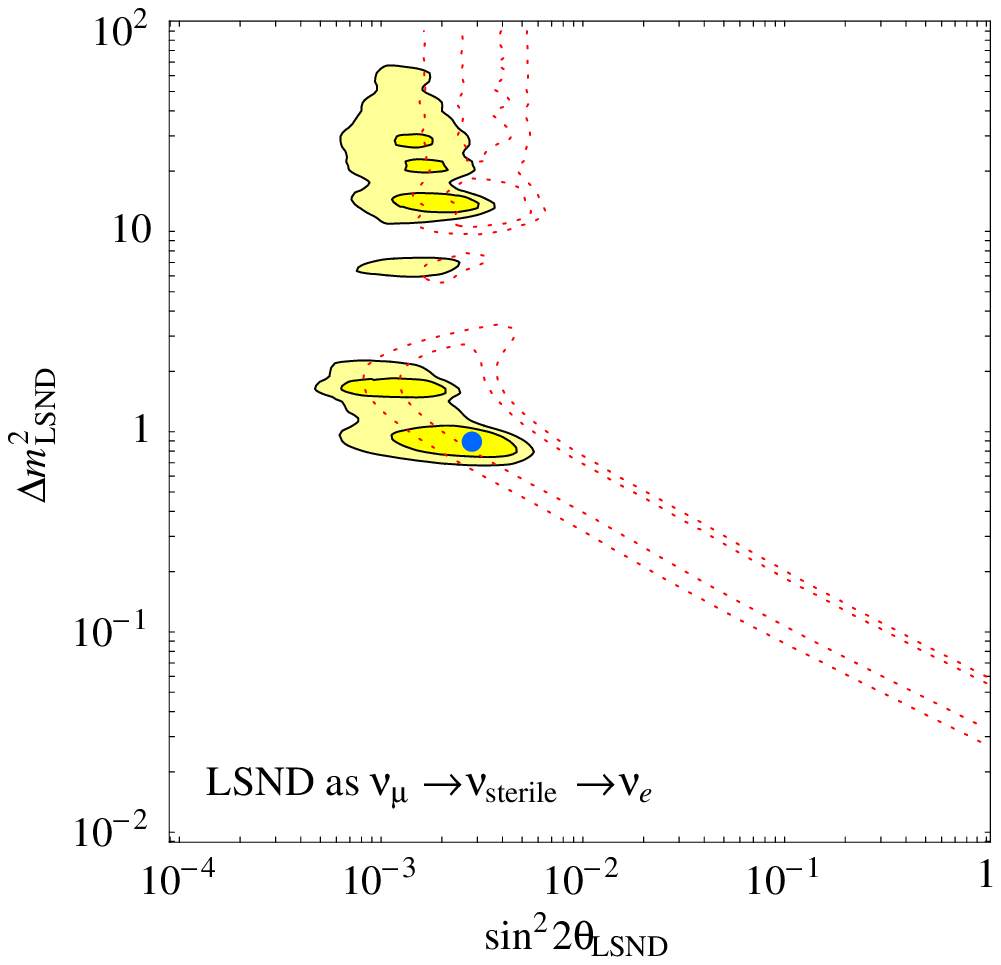}\hspace{1cm}
 \includegraphics[width=70mm]{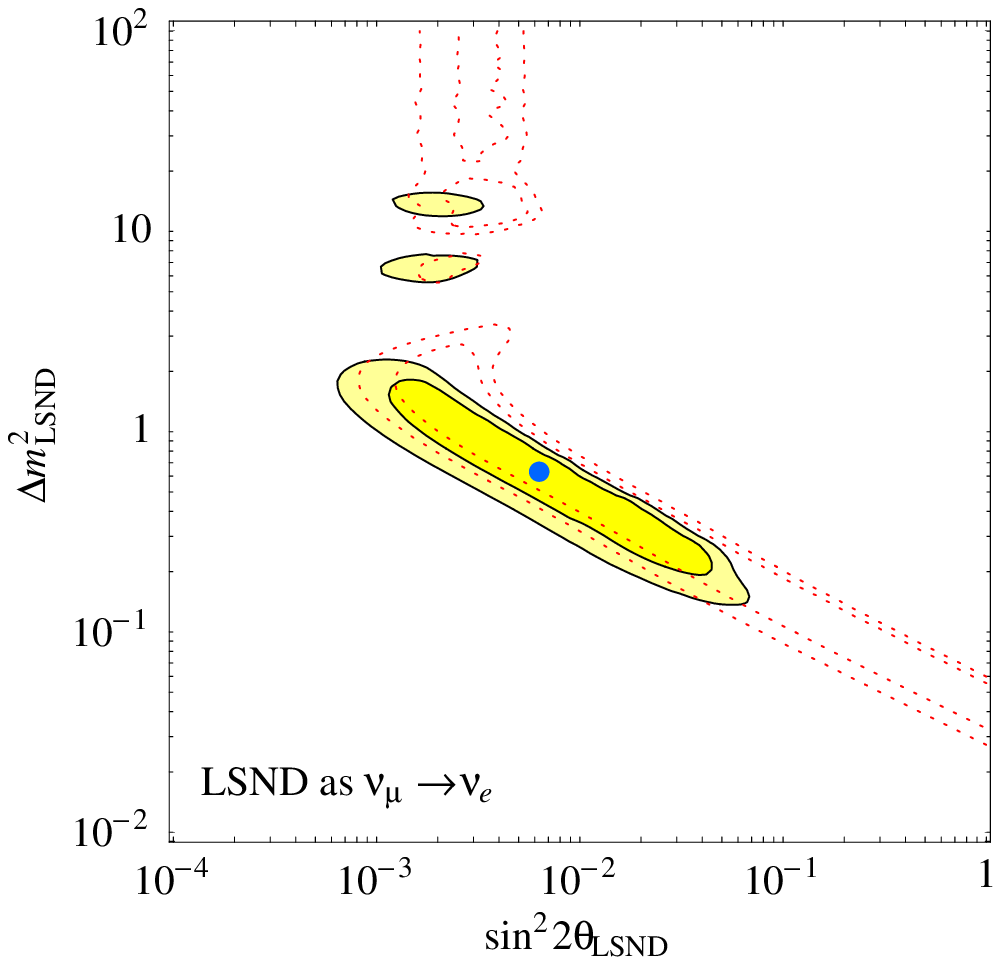}$$
  \caption[]{\em {\bf Best-fit regions at 90\% and 99\% CL (2 d.o.f.)
for the LSND parameters assuming oscillations.}
Fig.\fig{best}a assumes
that the LSND anomaly is generated trough a sterile neutrino
(`3+1' scheme).
Fig.\fig{best}b assumes that the LSND anomaly
is generated by active neutrinos, while something else 
(e.g.\ neutrino decay, sterile neutrinos,\ldots)
generates either the atmospheric or the solar anomaly,
without affecting LSND.
The dotted lines show the regions suggested by only the LSND data.
The dots show the best fit points.
\label{fig:best}}
\end{figure*}

The crucial question is if these bounds are too strong
for allowing the oscillations suggested by LSND.
At first sight the answer is that they are~\cite{3+1old},
but this negative conclusion was questioned in~\cite{3+1Smirnov}
and the first accurate statistical analysis of this issue was performed in~\cite{3+1first}
with Bayesian techniques.
Our result, shown in fig.\fig{split} basically agrees with~\cite{3+1first}.
Working in gaussian approximation\footnote{So that $\Delta\chi^2 = 7$ corresponds to
$97\%$ CL level for the two parameters $\theta_{\rm LSND}$ and $\Delta m^2_{\rm LSND}$.
The Gaussian approximation is not fully satisfied
(e.g.\ our best fit regions are not ellipses).
A Bayesian analysis can shift $97\%$ to $\sim 95\%$ or $\sim 98\%$, with `reasonable' choices
of the prior probability distribution.
(the arbitrarily remains until there are `large' allowed regions).
As discussed in~\cite{CSSSNO},
a similar shift is typically obtained in a frequentist analysis,
that cannot however be performed in a reasonable computing time.
Therefore we stick to the Gaussian approximation.} 
we find that all $96\%$ CL LSND confidence region is excluded
at, at least, $96\%$ CL level.
Therefore 3+1 solutions have some goodness-of-fit problem.
One needs to invoke a 
statistical fluctuation with around $\%$ probability to
explain why only LSND sees the sterile oscillations.

Even if this conclusion is self-evident,
we justify the adopted statistical strategy.
As discussed in~\cite{CSSSNO}, due to the large number of d.o.f.\
(about 200)
a na\"{\i}ve Pearson global $\chi^2$ test is unable
to notice this problem and would erroneously suggest that 3+1 oscillations give a good fit.
While it is difficult to develop a general and efficient good\-ness-of-fit test,
in this particular case the fit is bad for one specific reason: 
different sets of data are mutually exclusive (up to a $96\%$ CL)
within our theoretical assumptions.
In such a situation the goodness-of-fit problem is efficiently recognized by
fitting separately the two incompatible data.
This is what is done in fig.\fig{split}.

Ignoring the poor quality of the fit, the best combined fit region 
for the LSND parameters
is shown in
fig.\fig{best}a.
It agrees reasonably well with the corresponding fig.\ in~\cite{3+1last},
taking into account that we show values of 
$$\chi^2(\theta_{\rm LSND},\Delta m^2_{\rm LSND}) = \min_p \chi^2(p,\theta_{\rm LSND},\Delta m^2_{\rm
LSND})$$ 
(where $p$ are all other parameters in which we are not interested),
so that we convert values of $\chi^2-\chi^2_{\rm best}$ into confidence levels using the
gaussian values appropriate for 2 d.o.f.\ (the 2 LSND parameters),
while a statistically less efficient procedure with more d.o.f.\ is employed in~\cite{3+1last}.

\subsubsection*{2+2 neutrinos}
In the jargon 2+2 indicates 2 couples of
neutrinos (one generates the solar anomaly, and the other one
the atmospheric anomaly), separated by the large LSND mass gap.
Within this scheme, the sterile neutrino is employed to generate
the solar or atmospheric anomaly, or one combination of the two.
The fraction of sterile neutrino involved in solar oscillations, $\eta^{\rm sun}_{\rm s}$,
plus the fraction of sterile neutrino involved in atmospheric oscillations,
$\eta^{\rm atm}_{\rm s}$,
is predicted to sum to unity~\cite{3+1Smirnov}
$$ \eta^{\rm tot}_{\rm s}\equiv \eta^{\rm sun}_{\rm s} + \eta^{\rm atm}_{\rm s}=1.$$
Experiments now tell that
both the solar and atmospheric anomalies are mostly
generated by active neutrinos,
and only a small sterile contribution is allowed.
Consequently 2+2 oscillations give a global fit worse
than 3+1 oscillations~\cite{subMeV,3+1last}.
Let us summarize the present experimental status of this issue.


\begin{itemize}
\item {\bf Solar data} give a $5.4\sigma$ evidence for 
pure active solar oscillations versus pure sterile oscillations:
combining all solar data in a global fit we obtain~\cite{CSSSNO}
\footnote{Some words of caution. Arbitrary choices
become more relevant when fitting disfavoured data
(for example: the error is evaluated at the experimental point or at the theoretical point?).
Furthermore, our bound on the sterile fraction allowed by solar data 
is obtained assuming the BP00~\cite{BP00} prediction for the Boron $\nu_e$ solar flux.
It is proportional to the $^7{\rm Be}\,{\rm p}\to {}^8{}{\rm B}\, \gamma$
cross section: some authors think that systematic uncertainties in its measurement
 could be underestimated.}
$$\chi^2_{\rm sun}(\hbox{best sterile}) -\chi^2_{\rm sun}(\hbox{best active}) = 30 $$
and $\eta_{\rm s}^{\rm sun} = 0\pm 0.18$.
In particular, 
SNO/SK find a $5.1\sigma$ direct indication for $\nu_{\mu,\tau}$ appearance.

\item {\bf Atmospheric data}
 data give a $7\sigma$ indication for pure active atmospheric oscillations
versus pure sterile oscillations.
In fact, a global fit of atmospheric data gives~\cite{atmexp,SK02}\footnote{A large 
amount of these atmospheric data
is not included in theoretical reanalyses 
(because not yet accessible outside the SK collaboration 
in a form that allows to recompute them)
that therefore obtain a much smaller $\Delta \chi^2\approx 15$~\cite{3+1last,4nu}
in place of $50$~\cite{atmexp,SK02}.
This underestimation of the SK bound
means that {\em at the moment only SK can
perform a sensible analysis of mixed sterile and active 
atmospheric oscillations} and explains why the authors of~\cite{3+1last}
do not recognize that 2+2 oscillations are extremely disfavoured.
One mixing angle is set to zero in the SK analysis;
relaxing this unjustified simplification should not
significantly weaken the bounds.}
$$\chi^2_{\rm atm}(\hbox{best sterile}) -\chi^2_{\rm atm}(\hbox{best active}) \approx
50$$
and $\eta_{\rm s}^{\rm atm} = 0\pm 0.16$.
This strong evidence is obtained combining independent sets of data.
SK claims~\cite{atmexp} that pure sterile is disfavoured by 
the up/down ratio in a NC-enriched sample
($3.4$ standard deviations) and
by matter effects in partially contained events ($\approx2.9\sigma$) and
upward through-going muons ($\approx2.9\sigma$).
In total $7\sigma$~\cite{atmexp,SK02}.
Matter effects in MACRO~\cite{atmexp} give another $3.1\sigma$ signal.
Furthermore SK finds a direct $2\sigma$ hint for $\tau$ appearance.

\end{itemize}
In summary,
the two extreme cases
(all the sterile in atmospheric oscillations and
all the sterile in solar oscillations) have been excluded,
as summarized in table~\ref{tab:global}.
At the moment published results only allow
an approximated analysis of 
intermediate cases.
We find that
$\eta^{\rm tot}_{\rm s} = 0\pm 0.25$,
with $\eta^{\rm tot}_{\rm s}=1$ disfavoured at $4\sigma$.
Intermediate cases are less disfavoured than 
the two extreme cases
by only the amount expected, on a statistical basis, due to the presence of one more parameter:
the `best' fit is now obtained around 
the weighted average of the two incompatible solar and atmospheric determinations,
$\eta^{\rm atm}_{\rm s}=1-\eta^{\rm sun}_{\rm s}\approx 0.5$.
We do not present more precise results because fitting incompatible data makes little sense.
Despite the approximation, the final conclusion is clear:
2+2 oscillations are too strongly disfavoured to be considered as a viable possibly.

In fig.\fig{best}b we show the best-fit region for the LSND parameters,
assuming that the LSND anomaly is generated by oscillations of active neutrinos.
This result applies to a general class of models where
something different than oscillations between active neutrinos
is the source of the solar or atmospheric anomalies.
In particular it applies to 2+2 oscillations:
despite they are strongly disfavoured 
the LSND best-fit regions are unaffected by the problems with solar and atmospheric data,
and can therefore be reliably computed.


This region extends to values of the LSND parameters 
not accessible within 3+1 oscillations, see fig.\fig{best}.
Therefore the value of $P(\nu_\mu\to \nu_e)$ that will be measured at MiniBooNE
could discriminate between the two cases:
roughly, 3+1 oscillations prefer a value of $P(\nu_\mu\to \nu_e)$ somewhat smaller than
the one suggested by LSND.
Furthermore 3+1 spectra must be accompanied by a significant disappearance of $\nu_\mu$
at the LSND frequency.
For example, our 3+1 best-fit (marked with a dot in fig.\fig{best}a)
has $\sin^2 2\theta_{\mu s} = 0.2$, 
around the sensitivity of MiniBooNE.

Both 2+2 and 3+1 oscillation patterns can be realized with different neutrino spectra.
Since at the moment (and in the near future) no experiment can resolve the difference
we do not consider all possibilities.
For example, even knowing the oscillation parameters
and the type of spectrum, we could 
not safely predict
neutrinoless double $\beta$ decay signals.

\subsubsection*{Many sterile neutrinos}
As shown in the last paper in~\cite{3+1Smirnov}, many sterile neutrinos
cannot give a much better 3+1 fit than a single sterile neutrino.
Of particular interest are minimal models where right-handed neutrinos
live in a single extra dimension of radius $R$~\cite{Dim},
that could be identified with the LSND scale.
In such $3+\infty$ models the problematic prediction\eq{qqq} of 3+1 oscillations
becomes slightly
more problematic~\cite{KK}.
In fact, for small mixing angles and in the limit of averaged sterile
oscillations, we now have 
$\theta_{\rm LSND} \approx \sqrt{7/10}\,\theta_{es} \theta_{\mu s}$
in place of $\theta_{\rm LSND}\approx\theta_{es}\theta_{\mu s}$.
More importantly, the effective active/sterile mixing angles are now predicted to be
$$\theta_{\ell s}^2 = \frac{\pi^2}{3}|V_{\ell 3}|^2 \Delta m^2_{\rm atm} R^2$$
(for a hierarchical spectrum of active neutrinos, 
the other cases are more problematic).
The {\sc Chooz} bound on $V_{e3}$ 
(that will soon be tested and eventually strengthened 
by long-baseline experiments)
now gives
another constraint on $\theta_{es}$,
making this minimal model more problematic than 3+1 oscillations.
One can consider a large variety of less predictive non-minimal extra dimensional models.

In the case of sterile solar or atmospheric oscillations, 
many sterile neutrinos
can be less disfavoured that a single sterile neutrino.
As discussed above, pure atmospheric
sterile oscillations are 
disfavoured mostly by matter effects (in the earth), that
suppress $\nu_\mu\to \nu_s$ at large energy:
SK data are better fitted by $\nu_\mu\to \nu_\tau$ oscillations, 
unsuppressed by matter effects.
Even in the solar case, matter effects (in the sun)
contribute to determine how much SMA sterile oscillations are disfavoured~\cite{RR}.
In presence of a tower of many sterile neutrinos,
matter effects do not suppress sterile oscillations at large energy or density,
until there is a sufficiently heavy sterile resonance to cross.
However, sterile oscillations must be strongly matter suppressed
within a supernova. 
As discussed in~\cite{KK} supernov\ae{}
strongly constrain sterile towers that
continue up to masses of $10^{4\div 5}\eV$.
This is e.g.\ the case of an extra-dimensional Kaluza-Klein tower
that continues up to the TeV
scale~\cite{Dim}.
In conclusion, $(2+\hbox{many})$ oscillations
can be less disfavoured than $2+2$ oscillations.
However, even forgetting the lack of theoretical motivation,
it does not seem possible to achieve a really satisfactory fit.

\section{CPT violation}\label{CPT}

\subsubsection*{Theory}
The only safe result is that
CPT is conserved in Lorentz-invariant local quantum field theories (QFT).
Therefore CPT-violating effects can be obtained by abandoning locality or Lorentz
invariance:

\begin{itemize}
\item[1.] 
In local QFT, CPT violation can be induced if the Lorentz symmetry is broken,
 e.g.\ spontaneously by vacuum expectation values
of fields with spin 1 or higher,
or cosmologically by interactions with some `\ae{}ther',
or by a non-trivial extra-dimensional background, or...
\end{itemize}
This first possibility~\cite{Barger,Baren} seems not promising for LSND: like anomalous matter effects
and unlike oscillations, new effects
are not enhanced at low neutrino energy.
Therefore old experiments~\cite{fermilab} done at energies $2\div 3$ orders of magnitude higher than LSND,
disfavour the best fit Karmen/LSND region.
Furthermore in this context it seems difficult to
obtain $P_{ee}< 1/2$ (as suggested by the latest SNO data~\cite{sunexp}) in solar oscillations~\cite{deGouvea}.

Therefore we focus on the second possibility,
that could explain the LSND anomaly~\cite{MY}:
\begin{itemize}
\item[2.] 
Strings, branes,  quantum foams,
wormholes, non commutative geometry
(and other non local things like that)
suggest CPT-violating effects,
maybe suppres\-sed by
only one power of the quantum gravity scale $M$
(this case gives rise to interesting signals even for $M\sim 10^{19}\,{\rm GeV}$~\cite{boh}).

\end{itemize}
If an effect at that level were an unavoidable phenomenon, 
quantum gravity at the
TeV scale would be excluded by bounds
on the $K_0\bar{K}_0$ mass difference:
$$m_{K_0} - m_{\bar{K}_0} < 0.4~10^{-9} \eV.$$
The mass difference between neutrinos and anti-neutrinos
that could explain LSND is larger by many orders of magnitude:
we assume that CPT-violating effects are dominantly felt by neutrinos.

The generic Hamiltonian that describes
non relativistic systems (e.g.\ Kaons) violates CPT,
if the constraints from the underlying local relativistic QFT are not imposed.
In the case of relativistic systems (e.g.\ neutrinos)
one can mimic the standard Hamiltonian 
demanded
by local relativistic QFT 
(particles together with anti-particles)
but without imposing all the constraints
demanded by QFT (particles degenerate with anti-particles), so that
the generic Hamiltonian that describes 
free propagation of Dirac neutrinos
has different mass terms for $\nu$ and $\bar{\nu}$.
The social duty of studying how
CPT-violating neutrino masses can arise in popular fundamental models
has been exploited in~\cite{nulla},
obtaining the imprimatur from string brane-world orbifolds.
Non commutative geometry was invoked in~\cite{MY}.

We do not consider other possible CPT violations
in neutrino interactions, because 
experiments with (mainly) $\nu_\mu$, $\bar\nu_\mu$ beams
and precision electroweak data~\cite{LEPWWG,ellWnu}
find that neutrino NC couplings cannot differ from the SM prediction by
more than few $\%$.
A global fit of electroweak precision data~\cite{ellWnu} shows that
the CC couplings of $e$ and $\mu$ neutrinos
agree with the SM with few per-mille accuracy.

\subsubsection*{Fit of SK and K2K data}
In absence of oscillations, the number of $\nu_\mu$-induced
events at SK would be roughly double than the number of $\bar\nu_\mu$-induced events
(the ratio is higher at sub-GeV energies.
This is mainly due to the different $\nu_\mu$ and $\bar{\nu}_\mu$ cross-sections
on matter, 
that we
compute by summing the elastic
and deep-inelastic cross sections~\cite{Llewellyn}).
We assume that SK has an equal efficiency for
$\nu$ and $\bar\nu$-induced events.

We use a (hopefully) self-explanatory notation for the $\nu$ and $\bar{\nu}$
parameters.
An over-bar marks anti-neutrino parameters.
For example,  $\bar{\theta}_{\rm atm}$ and $\dmbatm$ parameterize the atmospheric
$\bar{\nu}_\mu\to \bar{\nu}_\tau$ oscillations.

\medskip

\paragraph{Restricted analysis}
To begin, we assume that $\theta_{\rm sun}$, 
${\theta}_{\rm CHOOZ}$,
$\bar{\theta}_{\rm CHOOZ}$, $\bar{\theta}_{\rm LSND}$
have negligible effect on atmospheric oscillations,
that are therefore described by
$\dmatm$, $\dmbatm$, 
$\theta_{\rm atm}$ and
$\bar{\theta}_{\rm atm}$.

\begin{figure*}[t]
$$\hspace{-10mm}
\includegraphics[width=80mm]{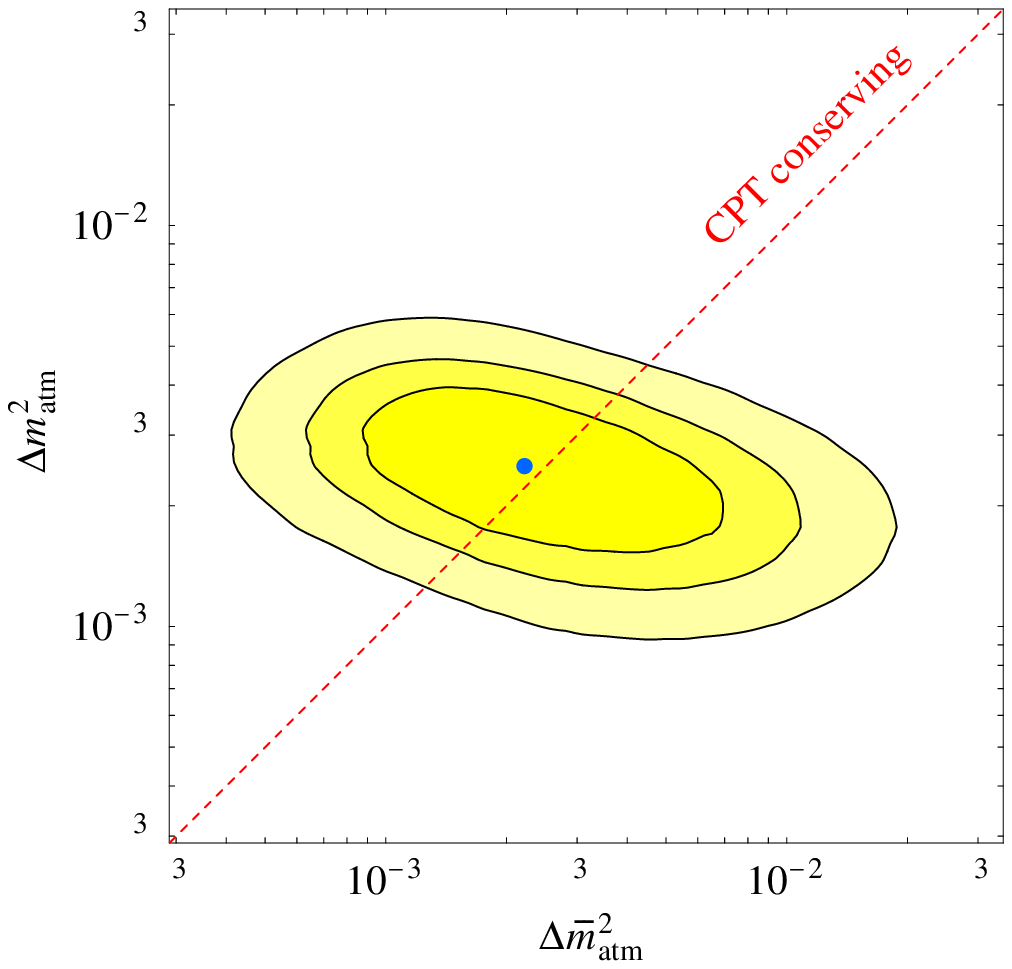} \hspace{14mm}
\includegraphics[width=80mm]{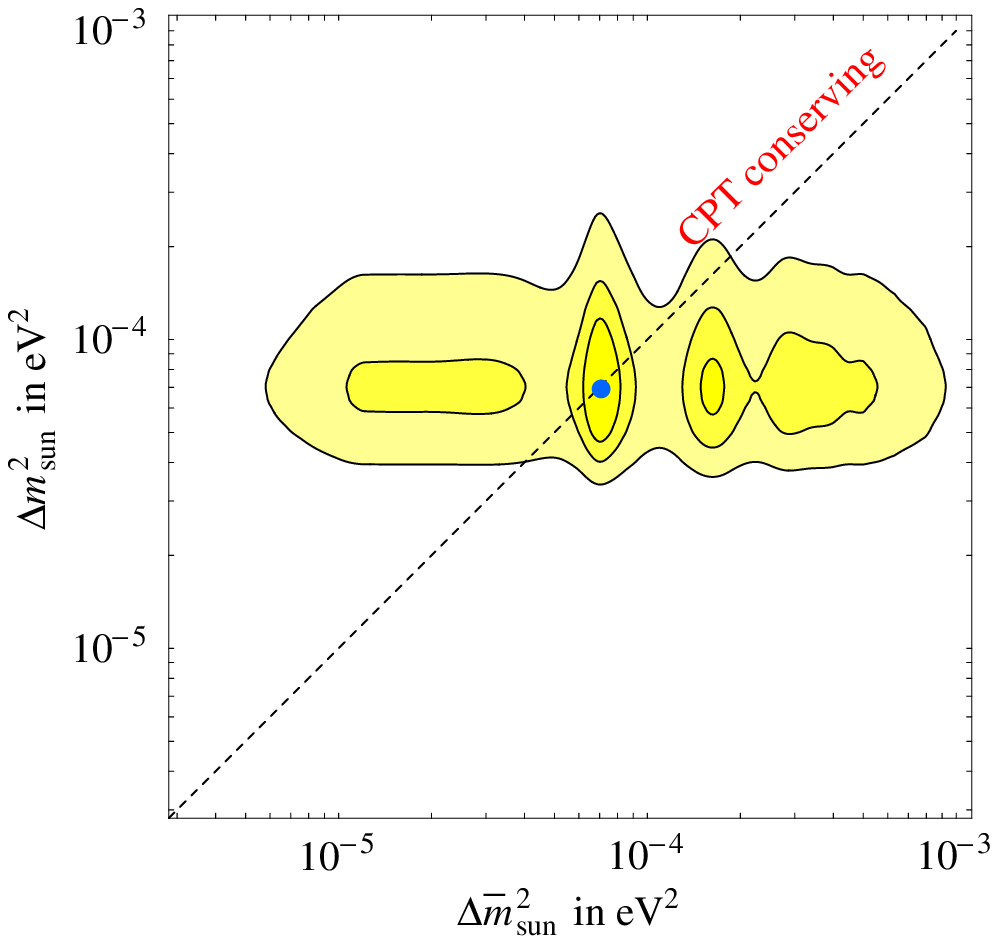}$$
\parbox{8.6cm}{  \caption[]{\em Fit of SK and K2K data for the neutrino and anti-neutrino
{\bf atmospheric} mass splitting at 68, 90, 99\% CL (2 d.o.f.).
\label{fig:CPTatm}}}\hspace{0.5cm}
\parbox{8.6cm}{  \caption[]{\em Fit of solar and reactor data for the neutrino and anti-neutrino
{\bf solar} mass splitting at 68, 90, 99\% CL (2 d.o.f.).
\label{fig:CPTsun}}}
\end{figure*}

A simple approximation captures the main properties of the fit.
The up/down asymmetry 
in the number of multi-GeV muon events is~\cite{atmexp}
$$A \equiv \frac{N_\downarrow - N_\uparrow}{N_\downarrow + N_\uparrow}=0.327 \pm 0.045$$
Assuming maximal mixings, in the CPT-conserving case one has
\begin{eqnsystem}{sys:A}
\dmbatm = \dmatm \approx 3~10^{-3}\eV^2: && A \approx 1/3\\[3mm]
\riga{The asymmetry is smaller in CPT-violating cases, e.g.}\\
\label{eq:1/4}
\dmbatm \gg \dmatm \approx 3~10^{-3}\eV^2: && A \approx 1/4\\
\label{eq:1/5}
\dmbatm \ll \dmatm \approx 3~10^{-3}\eV^2: && A \approx 1/5\\
\dmatm \gg \dmbatm \approx 3~10^{-3}\eV^2: && A \approx 1/7\\
\dmatm \ll \dmbatm \approx 3~10^{-3}\eV^2: && A \approx 1/11
\end{eqnsystem}
and even smaller if mixings are non maximal.
These considerations allow to understand the main features of our numerical
result.
In fig.\fig{CPTatm} we show the $\chi^2$ minimized with respect to the mixing angles
$\theta_{\rm atm}$ and $\bar\theta_{\rm atm}$.
While $\dmatm$ is almost as strongly constrained
as in a CPT-conserving fit, $\dmbatm$ can be about one order of magnitude larger or smaller
that $\dmatm$.\footnote{An analogous fit of sub- and multi-GeV SK data has
been performed in~\cite{svezia}, finding $\Delta\chi^2=\chi^2_{\hbox{\rm\tiny CPT}} - \chi^2_{\CPTV} = 16$,
while we do not find any strong evidence for CPT-violation.
As clearly discussed in~\cite{svezia}
this large $\Delta\chi^2$ could be an artifact due to having neglected the error 
on the ratio between $\nu_\mu$ and $\nu_e$ fluxes.
Our results also disagree with another CPT-violating fit presented in~\cite{Baren}:
the difference is significant even in the
CPT-conserving limit.
A fit performed by the SK collaboration~\cite{SK02} agrees with our fig.\fig{CPTatm}.

In the case of K2K data (sensitive to neutrinos)
we fitted the total number of events
ignoring the information about their energy,
finding a result in agreement with~\cite{lisiK2K}.}
The global $\chi^2$ for SK data is here obtained
by summing the $\chi^2$ corresponding to the individual
zenith-angle distributions of
sub-GeV and multi-GeV (10 $e$-like bins and 10 $\mu$-like bins each),
stopping $\mu$ (5 bins) and
upward-through-going $\mu$ (10 bins) events.
The overall normalization in each kind of events has been considered as a free parameter.

Alternatively, one can try to take into account the theoretical
predictions for the overall fluxes as in~\cite{Lisi} employing a $55\times 55$ correlation matrix.
This second approach gives a slightly different bound on CPT-violation:
larger values of $\dmbatm$ would not be significantly disfavoured up to the right border of fig.\fig{CPTatm}.


Since the best fit is obtained for almost CPT-conserving oscillations,
the fit for the mixing angles is quite simple,
and we do not need to show a dedicated figure.
In the CPT-conserving case
$\sin^2 2\theta_{\rm atm}$ has to be close to one.
We find that in the CPT-violating case the same bound applies replacing
$$\sin^2 2\theta_{\rm atm}\to
\frac{2}{3}\sin^2 2\theta_{\rm atm}+\frac{1}{3}\sin^2 2\bar{\theta}_{\rm atm}$$
so that both $\theta_{\rm atm}$ and (to a lesser extent) $\bar{\theta}_{\rm atm}$
have to be close to maximal.

\paragraph{General analysis}
We now discuss the effects of the other mixing angles,
$\theta_{\rm sun}$, 
${\theta}_{\rm CHOOZ}$,
$\bar{\theta}_{\rm CHOOZ}$, $\bar{\theta}_{\rm LSND}$, that we have so far neglected.
Some of them are allowed to be large,
but cannot significantly affect our CPT-violating atmospheric fit
shown in fig.\fig{CPTatm}.

\setlength{\unitlength}{1mm}
\begin{table*}
$$\begin{array}{lc|c|cccc}
\multicolumn{2}{c|}{\hbox{model and number of free parameters}} & \Delta\chi^2_{\rm tot}
&\Delta \chi^2_{\rm sun} & \Delta\chi^2_{\rm atm} & \Delta\chi^2_{\rm LSND}& \Delta\chi^2_{\rm bounds} \\ \hline
\hbox{3 neutrinos and \begin{picture}(12,0)(0,0)
  \put(0,0){CPT}\put(0,0){\line(3,1){8}}\end{picture}}& 10 &  0\hbox{ (best fit)}  & 0      & 0 & 0.5   & 3.2 \\
3+1:~\Delta m^2_{\rm sterile} = \Delta m^2_{\rm LSND} & 9  & 6   & 0      & 0 & 2.9   & 6.7   \\
\hbox{normal 3 neutrinos}                             & 5  & 25  & 0      & 0 & 28.8  & 0     \\
2+2:~\Delta m^2_{\rm sterile} = \Delta m^2_{\rm sun}  & 9  & 30  & 30     &0  & 0.5   & 3.2   \\ 
2+2:~\Delta m^2_{\rm sterile} = \Delta m^2_{\rm atm}  & 9  & 50  & 0      &50 & 0.5   & 3.2   
\end{array}$$ 
\caption{\em {\bf Interpretations of all oscillation data, ordered according to
the quality of their global fit.}
The last 4 columns show the minimal $\Delta\chi^2$ restricted
to solar data, atmospheric data, LSND data, and to
experiments compatible with no oscillations (mainly {\sc Chooz}, {\sc Bugey} and CDHS).\label{tab:global}}
\end{table*}

In anti-neutrinos, disappearance experiments require
small values of the two mixing angles that 
induce oscillations at the LSND frequency.
These constraints allow for a novel possibility,
somewhat disfavoured only by atmospheric data:
the most splitted anti-neutrino eigenstate could
be dominantly $\bar{\nu}_\mu$ (rather than $\bar{\nu}_e$ as in fig.\fig{CPTspettro}).
In this case, $\bar\theta_{\rm CHOOZ}$
(the remaining mixing angle that now gives oscillations at the atmospheric frequency)
could be large, without conflicting with the {\sc Chooz} bound, if
$\dmbatm$ is below the {\sc Chooz} sensitivity.

In neutrinos, solar experiments require
$\theta_{\rm sun}\sim 1$ as in the CPT-conserving case.
Unlike in the CPT-conserving case {\sc Chooz} does not force
$\Delta m^2_{\rm sun}\circa{<}0.7~10^{-3}\eV^2$,
but a larger $\Delta m^2_{\rm sun}$ has recently been disfavoured by the SNO NC data~\cite{sunexp}.
The angle $\theta_{\rm CHOOZ}$ (that induces $\nu_\mu\to \nu_e$ oscillations at the
atmospheric frequency; we improperly adopt the name used in CPT-conserving analyses)
is not bounded by {\sc Chooz} (i.e.\ by disappearance of $\bar{\nu}_e$),
but only by global fits of solar and atmospheric data,
that weakly disfavour a large $\theta_{\rm CHOOZ}$~\cite{global3nu}.

\paragraph{Signals}
At the light of these results, we can now list the CPT-violating signals
that could appear in forthcoming experiments (some signals were discussed in~\cite{MY,nulla,Baren})

\begin{itemize}

\item MiniBooNE will not see the LSND oscillations,
if will only search them as $\nu_\mu\to\nu_e$ rather than as $\bar{\nu}_\mu\to\bar{\nu}_e$.
\end{itemize}
While this signal is mandatory if 
the CPT-violating interpretation of the LSND anomaly is correct,
the following signals can but need not to appear, depending on the values
of the unknown parameters:
\begin{itemize}

\item We would have a signal for CPT violation if
KamLAND will find no solar oscillations in its reactor data, and Borexino
will indirectly favour LMA by finding a $\sim 1/2$ suppression and
no matter nor seasonal effects.

\item If $\bar{\theta}_{\rm CHOOZ}$ were large, KamLAND would discover its effects
and misinterpret them as LMA oscillations.
In particular this means that if KamLAND will confirm LMA, 
a CPT-violating interpretation of the LSND anomaly would not be immediately excluded,
but only disfavoured.
We do not list other possible situations that could happen
depending on future Borexino and KamLAND results.
\footnote{When these results will be announced,
we will update the hep-ph version of this paper, adding
a precise discussion.}
\label{CPTKL}

\item According to our fit in fig.\fig{CPTatm},
long-baseline experiments that plan to employ a $\nu_\mu$ beam
(like K2K, Minos and CNGS)
have almost the same capabilities
of confirming atmospheric oscillations as in the CPT-conserving case.
Using a a  $\bar{\nu}_\mu$ beam they can also
test if $\dmbatm$ is higher than $\dmatm$ 
(if $\dmbatm$ is as large as possible, a $5\%$ $\bar{\nu}_\mu$ contamination
in the $\nu_\mu$ beam could also give detectable $\tau$-appearance effects).

\item These long-baseline experiments
can test if
$\theta_{\rm CHOOZ}$ 
is larger than what allowed in the CPT-conserving case 
by looking at $\nu_\mu\to \nu_e$.


\end{itemize}
In longer terms, an atmospheric experiment
that separately measures $\dmatm$ and $\dmbatm$
(and sees the first oscillation dip) seems feasible~\cite{monolith},
although KEK, CERN and FermiLab
preferred to pursue 3 long-baseline experiments.

With a hierarchical $\bar{\nu}$ spectrum
(rather than with the inverted spectrum motivated in~\cite{MY})
planned $\beta$-decay experiments
like KATRIN~\cite{KATRIN}
can test the upper part of the $\Delta m^2$ range suggested by LSND~\cite{nulla}.
Planned neutrino-less double $\beta$-decay experiments~\cite{Klapdor}
have brighter perspectives of improvement than $\beta$-decay experiments,
but CPT-violating neutrino masses seem to require
Dirac (rather than Majorana) neutrinos,
if the Lorentz symmetry is unbroken
(because there is no Lorentz-invariant distinction between
massive Majorana $\nu$ into a $\bar{\nu}$:
a sufficiently `fast' Lorentz transformation transforms $\nu$ in $\bar{\nu}$).

In the far future, with a neutrino factory it should be possible to
test CPT conservation in atmospheric oscillations
at the $\%$ level~\cite{CPTnuf}.


\section{Conclusions}
A possible global explanation of the three neutrino anomalies (atmospheric, solar and LSND)
is that an extra sterile neutrino generates one of them.
Each anomaly, when fitted independently from the other ones, 
prefers active oscillations refusing the sterile neutrino.
The relatively better global fit is obtained
with a 3+1 spectrum (sterile LSND oscillations)
rather than with a 2+2 spectrum (sterile solar or atmospheric oscillations:
this case is disfavoured at $4\sigma$, after the recent SNO NC results~\cite{sunexp}).
However the fit is not good: within the 3+1 scheme the LSND anomaly conflicts with
$\nu_e$ or $\nu_\mu$ disappearance experiments.
One needs to invoke a statistical fluctuation with around $\%$ probability
to understand why {\sc Bugey}, {\sc Chooz}, CDHS or SK have not seen sterile effects.
Our main results are summarized in table~1.


The best-fit LSND regions are shown in fig.\fig{best},
assuming that the LSND anomaly is generated trough a sterile neutrino (3+1 case, fig.\fig{best}a)
or by oscillations of active neutrinos (fig.\fig{best}b),
assuming that a sterile neutrino or something else (e.g.\ neutrino decay)
generates the solar or atmospheric anomaly.
The best fit LSND regions are somewhat different:
MiniBooNE could discriminate the two cases.

Many sterile neutrinos (motivated e.g.\ in extra dimensional models)
can somewhat improve the fit,
but it does not seem possible to obtain a good sterile solution.

\medskip

In view of these unsatisfactory sterile fits, and
of the latest LSND results~\cite{LSND}
\begin{eqnarray*}
P(\nu_\mu\to \nu_e) &=&  (1.0  \pm 1.6) \, 10^{-3}\\
P(\bar\nu_\mu\to \bar\nu_e) &=&   (2.6 \pm 0.8)\, 10^{-3}
\end{eqnarray*}
one might want to speculate on CPT-violation.
A satisfactory global fit of all neutrino data (see table~1)
can be obtained with the CPT-violating neutrino masses proposed in~\cite{MY}.
Theory gives no useful restriction, and in particular does not tell
if CPT should be violated also in atmospheric oscillations,
although it looks plausible.
Fig.~\ref{CPT} shows how present
SK and K2K data restrict the atmospheric oscillation parameters
$\dmatm$ and $\dmbatm$.
They can differ by about one order of magnitude.
In section~2 we studied which CPT-violating oscillations
are compatible with present data, and listed the unusual signals
that could be seen at forthcoming solar (KamLAND, Borexino) and long-baseline experiments
(K2K, MINOS, CNGS) --- and of course at MiniBooNE.


%
%

\paragraph{Acknowledgments}
We thank K. Eitel and M. Steidl
for giving us the table of the {\sc Karmen} likelihood,
and B. Louis and
G. Mills for the LSND likelihood.
We thank V. Berezinsky, R. Rattazzi and F. Vissani for useful discussions.


\def\baselinestretch{0.90}

\def\baselinestretch{1.27}\small\normalsize

 \setcounter{footnote}{1}

\section{Addendum about the first KamLAND and WMAP data} \label{in}

\paragraph{CPT-violating solution}
The CPT-violating neutrino spectrum suggested in~\cite{MY, nulla} allowed to reconcile the
solar, atmospheric and LSND neutrino anomalies
but predicted no effect in KamLAND.
The evidence seen by KamLAND~\cite{KamLAND} can be fitted by
the alternative CPT-violating spectrum proposed at page~\pageref{CPTKL} of the original version of this paper,
at the price of a non standard fit of atmospheric data.
We now show that, as anticipated in~\cite{oursunfit}, this solution is disfavoured by atmospheric data
(we disagree with the claim in~\cite{BBL} that it gives an atmospheric fit
``clearly favored over the CPT conserving one'').

We denote anti-neutrino parameters with an over-bar.
Fitting SK and K2K data in the usual two neutrino approximation,
in fig.\fig{CPTatm} we plotted the best-fit regions for the atmospheric mass splittings
($\Delta \bar{m}^2_{\rm atm}$,
$\Delta m^2_{\rm atm}$), marginalizing the global $\chi^2$ with respect to
the atmospheric  mixing angles $\theta_{\rm atm}$ and $\bar{\theta}_{\rm atm}$.
We can now do the same test on solar $\nu$ data and reactor $\bar\nu$ data
(from the Homestake, SAGE, Gallex, GNO, SK, SNO, KamLAND, CHOOZ experiments).
The result is shown in fig.\fig{CPTsun}.
The plot is restricted to the LMA solution:
other solutions with smaller $\Delta m^2_{\rm sun}$ (LOW and QVO solutions)
are disfavoured but not excluded by solar $\nu$ data.
The best fit is close to CPT-conservation in both the atmospheric and in the solar cases.
We do not show the corresponding fits for the mixing angles
$\theta_{\rm atm}$, $\bar{\theta}_{\rm atm}$, $\theta_{\rm sun}$ and $\bar{\theta}_{\rm sun}$.
All these mixing angles must be large.

After including KamLAND data,
CPT-violation can  no longer perfectly fit all data.
We now have some evidence of `solar' and `atmospheric' oscillations
not only in $\nu$ but also in $\bar\nu$,
leaving no room for the larger $\Delta m^2$ that should give rise to the LSND anomaly.

In both `solar' and `atmospheric' cases $\bar\nu$ data do not yet provide a conclusive evidence,
so that we may attempt to fit all neutrino anomalies by sacrificing either 
i) solar $\bar\nu$ data
or ii) atmospheric $\bar\nu$ data.

Concerning case i), we just mention that the CPT-violating spectrum
proposed in~\cite{MY,nulla}
(that predicted no ano\-ma\-ly in KamLAND)
still gives a reasonably good global fit.
We do not consider this possibility,
that KamLAND should exclude with more statistics.

Rather, we explore case ii)
and sacrifice atmospheric data, taking
the smaller anti-neutrino $\Delta \bar{m}^2$ in the KamLAND (KL) range,
rather than in the atmospheric range,
and the larger $\Delta \bar{m}^2$ in the LSND range.
There are three anti-neutrino mixing angles.
Two of them (that from now on we name $\bar{\theta}_{\rm LSND}$ and $\bar{\theta}_{\rm atm}$)
give oscillations at the larger LSND frequency, $\Delta m^2_{\rm LSND}$.
The third mixing angle
(that from now on we name $\bar \theta_{\rm KL}$) gives oscillations at the smaller $\Delta m^2_{\rm KL}$.
KamLAND and LSND data want a large $\bar{\theta}_{\rm KL}\sim 1$
and a small $\bar{\theta}_{\rm LSND}\sim (0.1\div 0.01)$.
The last mixing angle, $\bar{\theta}_{\rm atm}$, induces
$\bar{\nu}_\mu \leftrightarrow\bar{\nu}_\tau$ oscillations at the LSND frequency.
As usual, the relatively better atmospheric fit is obtained for maximal $\bar{\theta}_{\rm atm}$.


Beyond performing a global fit, it is useful to present
a semi-quantitative understanding of SK data.
For our purposes the main observables are 
the number of up-ward going and down-ward going
$\mu$-like events in the multi-GeV sample.
We recall that SK cannot distinguish $\nu$ from $\bar\nu$.
Neutrinos have {\em roughly} the same flux and a two times larger cross section than anti-neutrinos:
Analytical estimates are performed by just using this factor two.
The total flux has a large overall uncertainty.

We first give an argument similar to the one used in~\cite{BBL}
to state that atmospheric data favour CPT-violation,
so that a comparison shows the reason of the disagreement.
\begin{itemize}
 \item[1.]  The maximal up/down asymmetry
that CPT-conser\-ving
$\nu_\mu\to\nu_\tau$ oscillations
can produce in the `multi-GeV $\mu$-like + PC' sample
of SK atmospheric data is (eq.\eq{1/4})
$$A_{\rm ideal} \equiv \frac{N_\mu(\cos\vartheta = 1) - N_\mu(\cos\vartheta = -1)}{N_\mu(\cos\vartheta = 1) + N_\mu(\cos\vartheta = -1)}=\frac{1}{3}$$
where $\vartheta$ is the zenith angle and
$\cos\vartheta = 1$ corresponds to vertical down-going events.
The corresponding upward/downward asymmetry is
$$A_{\rm real}\equiv \frac{N_\mu(\cos\vartheta >0.2) - N_\mu(\cos\vartheta <-0.2)}{N_\mu(\cos\vartheta >0.2) + N_\mu(\cos\vartheta <-0.2)}= 0.28,$$
defined ignoring `horizontal' events with
$|\cos \theta|< 0.2$.
The most recent SK data~\cite{SK2K} give the value 
$$A_{\rm real}=0.288\pm0.030.$$
The proposed CPT-violating scenario
can give at most an up/down asymmetry $A_{\rm ideal} = 1/4$,
which corresponds to $A_{\rm real}=0.21$,
$2.5\sigma$ below the experimental value.
\end{itemize}
This argument takes into account only a part of SK data;
furthermore the precise value of $A_{\rm real}/A_{\rm ideal}$ depends on $\Delta m^2_{\rm atm}$.
We present another argument which avoids these drawbacks:
\begin{itemize}
\item[2.] Since $\Delta m^2_{\rm LSND}\gg \Delta m^2_{\rm atm}$ the
 proposed CPT-violating oscillations give a reduction in the muon rate with an energy and
zenith-angle dependence which
(up to an overall factor that does not play an important r\^ole in the SK analysis)
can be mimicked by normal CPT-conserving oscillations
with an appropriate effective value of the mixing angle
$$  \underbrace{\sin^2 2\theta_{\rm atm}}_{\rm CPT-conserving}\leftrightarrow 
\underbrace{\frac{4\sin^2 2\theta_{\rm atm}}{6-\sin^2 2\bar\theta_{\rm atm}}}_{\Delta \bar m^2_{\rm atm}\gg \Delta m^2_{\rm atm}}\le \frac45.$$
The maximal value of the effective CPT-conserving $\sin^2 2\theta_{\rm atm}$
allowed by the proposed CPT-violating oscillations is $4/5$, which is
 $5\sigma$ below
the experimental value
$$\sin^2 2\theta_{\rm atm} =1.00 \pm 0.04~\cite{SK2K}.$$
\end{itemize}
Our global fit of SK data, shown in fig.\fig{CPTatm}, gives
$$\chi^2 (\Delta\bar{m}^2_{\rm atm} = \Delta m^2_{\rm LSND}) - 
\chi^2 (\Delta\bar{m}^2_{\rm atm} = \Delta m^2_{\rm atm}) \approx 5^2$$
confirming the second argument.
We remark that we fit all SK data and not only the up/down asymmetries.
We computed the $\Delta\chi^2$ using the latest SK data (1489 days of data taking\footnote{We thank M. Shiozawa for providing us the data presented at the Neutrino 2002 conference.
Using the slightly older SK data set employed in the original version of this paper
would make no significant difference.}~\cite{SK2K}).
Our numerical code contains precise neutrino and anti-neutrino fluxes and cross-sections.

\begin{figure*}[t]
$$
\includegraphics[width=70mm]{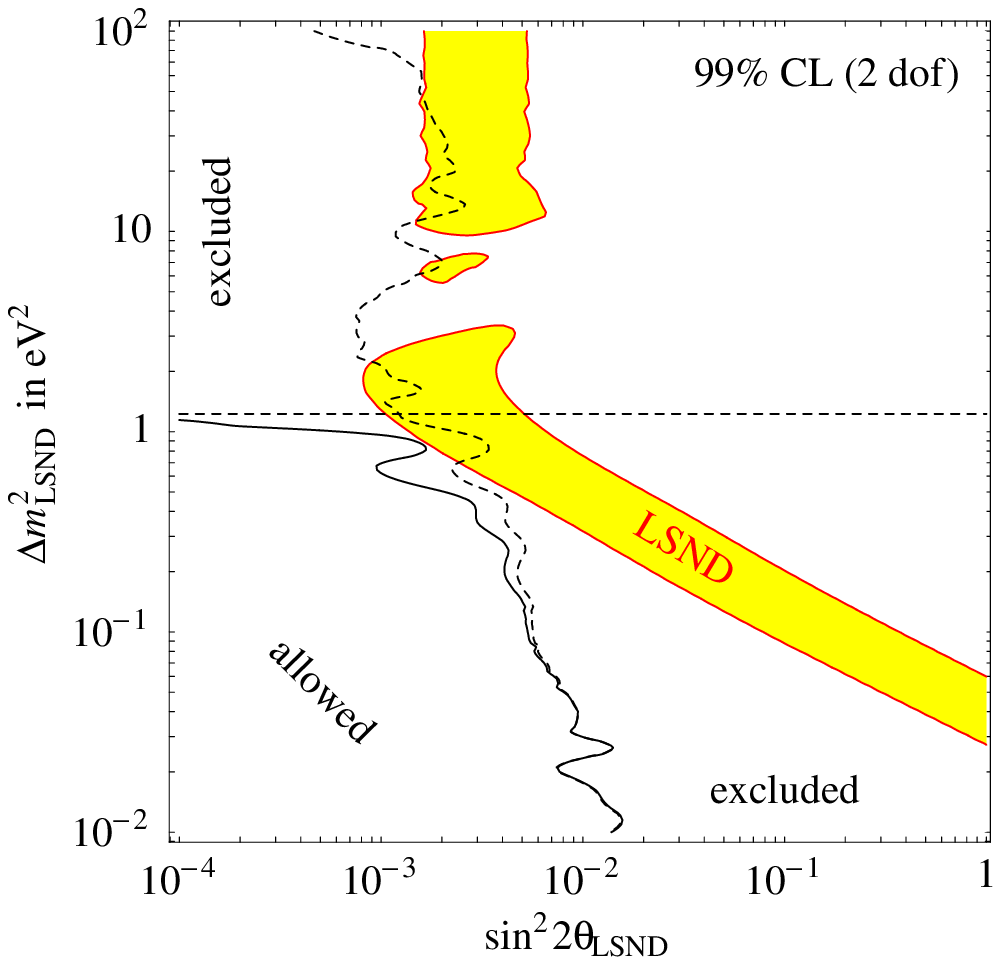} \hspace{1cm}
\includegraphics[width=70mm]{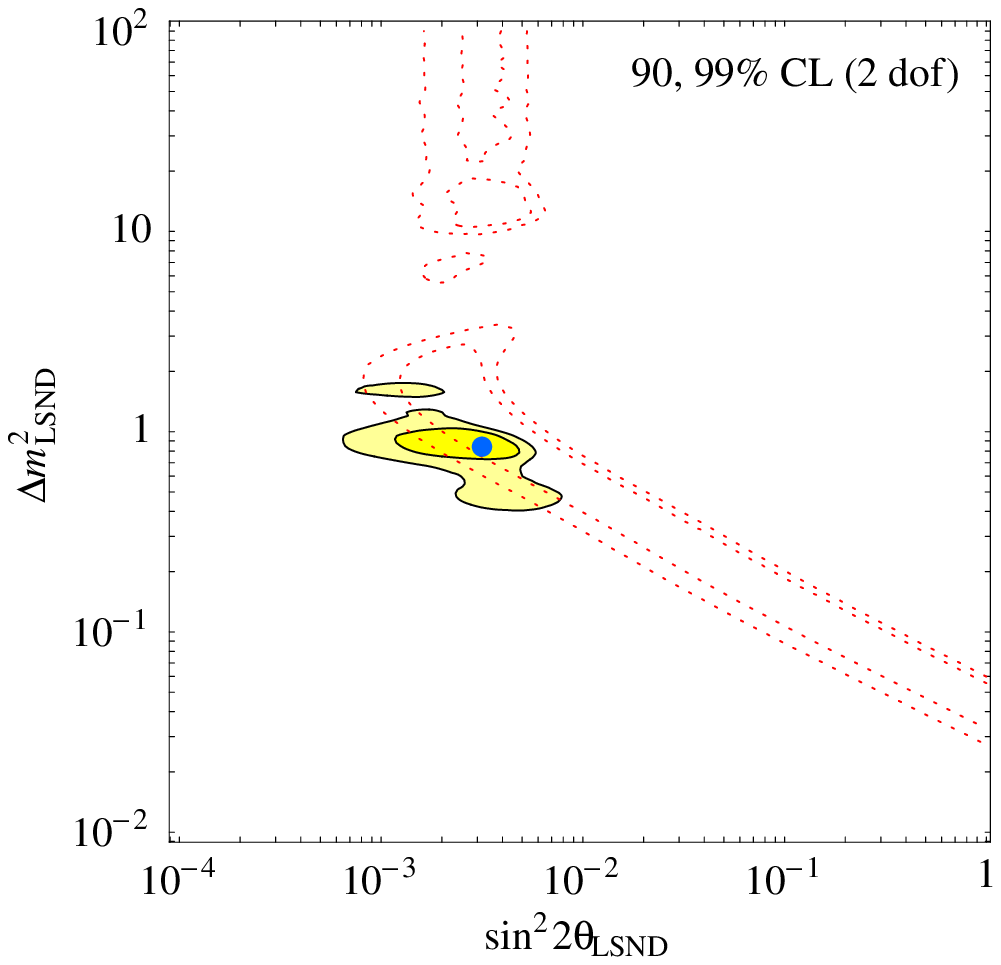}$$
\caption[]{\em {\bf Status of 3+1 oscillations including WMAP data}.
Fig.\fig{31MAP}a: LSND favours the shaded region.
Values of $\Delta m^2_{\rm LSND}$ above the horizontal dashed line are disfavoured by WMAP.
The other dashed line shows the upper bound on $\theta$ from all other neutrino experiments.
The continuous line shows the combination of the two previous constraints.
{\em All} bounds are at $99\%$ CL for 2 dof.
In fig.\fig{31MAP}b we show the best fit 3+1 solution, including all data.
\label{fig:31MAP}}
\end{figure*}



\paragraph{$L$-violating muon decay}
We extend our analysis considering one more 
tentative interpretation of the LSND anomaly, in terms of a
speculative $\Delta L = 2$ muon decay channel
with branching ratio roughly equal to 
the oscillation probability suggested by LSND:
$$\hbox{BR}(\bar\mu\to\bar e\bar\nu_e\bar\nu)\approx P(\bar\nu_\mu\to\bar\nu_e)=(2.6\pm0.8)~10^{-3}.$$
Since the fact that LSND has a longer path-length than {\sc Karmen} plays no r\^ole
according to this interpretation,
one na\"{\i}vely expects that it is disfavoured by {\sc Karmen} 
as much as oscillations with large $\Delta m^2$.

This expectation was questioned in~\cite{Babu}, that presented one explicit model
that produces the $\bar\mu\to\bar e\bar\nu_e\bar\nu$ decay
with Michel parameter $\rho =0$ (while  $\rho = 3/4$ in ordinary muon decay) and consequently
a $\bar\nu_e$ spectrum softer than the $\bar\nu_e$ spectrum produced by oscillations with large $\Delta m^2$.
As a consequence the {\sc Karmen} bound on $\bar\nu_e$ appearance
gets relaxed~\cite{Babu}  by a factor $\lambda =1.9$~\cite{KarmenLast}\footnote{We thank Klaus Eitel for communications about the {\sc Karmen} analysis.} with respect to the bound obtained from the
 analysis in terms of oscillations with large $\Delta m^2$.
In fact, {\sc Karmen} detects $\bar\nu_e$ using the
$\bar\nu_e p\to \bar{e} n$ reaction, which cross-section is roughly proportional to 
$E_{\bar\nu_e}^2$.

However also the LSND experiment detects $\bar\nu_e$ using the
$\bar\nu_e p\to \bar{e} n$ reaction.
Therefore an interpretation of the LSND anomaly needs a $\hbox{BR}(\bar\mu\to\bar e\bar\nu_e\bar\nu)$ larger than
$P(\bar\nu_\mu\to\bar\nu_e)$ by the same factor $\lambda$.
The two correction factors compensate each other when comparing
LSND with Karmen, indicating that the na\"{\i}ve expectation is right.
Table~\ref{tab:global2} quantifies how much the $\bar\mu\to\bar e\bar\nu_e\bar\nu$ solution is disfavoured.
A fully precise result would need a dedicated analysis,
that only the LSND collaboration can perform.

\setlength{\unitlength}{1mm}
\begin{table*}[t]
$$\begin{array}{lc|ccc}
\multicolumn{2}{c|}{\hbox{model and number of free parameters}} & \Delta\chi^2 & \hbox{mainly  incompatible with} \\ \hline
\multicolumn{2}{c|}{\hbox{ideal fit (no known model)}}& 0 &\\
\Delta L = 2\hbox{ decay }\bar\mu\to\bar e \bar\nu_\mu\bar\nu_e & 6 &  12 &\hbox{Karmen} \\
3+1:~\Delta m^2_{\rm sterile} = \Delta m^2_{\rm LSND} & 9  & 15     & \hbox{Bugey, WMAP} \\
\hbox{3 neutrinos and \CPTVbig (no $\Delta \bar m^2_{\rm sun}$)}& 10 &  15  & \hbox{KamLAND} \\
\hbox{3 neutrinos and \CPTVbig (no $\Delta \bar m^2_{\rm atm}$)}& 10 &  25  & \hbox{SK atmospheric} \\
\hbox{normal 3 neutrinos}                             & 5  & 25        & \hbox{LSND} \\
2+2:~\Delta m^2_{\rm sterile} = \Delta m^2_{\rm sun}  & 9  & 30     & \hbox{SNO} \\
2+2:~\Delta m^2_{\rm sterile} = \Delta m^2_{\rm atm}  & 9  & 50    & \hbox{SK atmospheric} \\
\end{array}$$ 
\caption{\em {\bf Interpretations of all oscillation data, ordered according to
the quality of their global fit}.
A $\Delta \chi^2 = n^2$ roughly signals an incompatibility at $n$ standard deviations.
\label{tab:global2}}
\end{table*}

\paragraph{2+2 sterile neutrinos}
Ref.s~\cite{Foot,thetini} questioned the conclusion that an interpretation of the LSND anomaly
in terms of an extra sterile neutrino with `2+2' spectrum
have been excluded by solar and atmospheric experiments.

The author of~\cite{Foot} estimates that 2+2 oscillations provide a global fit of all neutrino data with
$\chi^2/{\rm dof} \approx 291/276$,
which is acceptable.
This is true, but the goodness-of-fit (gof) test based on the value of the total $\chi^2$ is inefficient when ${\rm dof}\gg 1$:
it may assign an acceptable gof probability to a solution which is already excluded.
This issue was discussed in the context of analyses of solar data in~\cite{oursunfit}
and can be exemplified by recalling that, according to CPT-conserving global fits of solar and KamLAND data~(e.g.~\cite{oursunfit}),
the LOW solution have been excluded but its na\"{\i}ve gof is still acceptable (presently it has 
$\chi^2/{\rm dof} = 89/91$).

The authors of~\cite{thetini} pointed out that the full $4\times 4$ mixing matrix contains
small mixing angles which could relax the solar and atmospheric constraints
on 2+2 oscillations.
We do not expect that these effects can significantly improve the 2+2 status.

\paragraph{3+1 sterile neutrinos}
After the first WMAP data cosmology gives the dominant bound on neutrino masses~\cite{WMAP}
$$\sum m_\nu <0.69\eV\qquad\hbox{(95\% CL, 1 dof)}.$$
We assume that the extra sterile neutrino has a thermal abundancy
(a possibility still compatible with primordial nucleosynthesis,
unless uncertainties are aggressively estimated~\cite{BBN}),
so that its mass must be included in the sum.
Both the WMAP bound and the assumed sterile abundancy are valid in standard cosmological model
but might not hold in general.
Present data are consistent with the standard cosmological model, and disfavour
reasonable concrete mechanisms of evading the bound on $\sum m_\nu$
(for recent discussions see e.g.\ \cite{BBN, Mura}).
Therefore, keeping in mind the possible caveat that the 
bound on $\sum m_\nu$ could be weaker in some ad hoc cosmological  model,
we add this bound to our data-set.
Extracting the precise probability distribution for $\sum m_\nu$ from~\cite{WMAP},
we study its consequences for LSND.

As already remarked in~\cite{Mura,Giunti}
(which performed semi-quantitative analyses and do not fully agree among them),
the WMAP bound disfavours the 3+1 solution, 
since it employs a sterile neutrino with mass $(\Delta m^2_{\rm LSND})^{1/2}\simeq \sum m_\nu$.
The precise combined bound is shown in fig.s\fig{31MAP}, explained in their captions.
These figures update our older fig.s\fig{split} and\fig{best}a,
and section~\ref{sterile} describes in greater detail how they are obtained.
Table~\ref{tab:global2} quantifies how much the 3+1 solution is now disfavoured.

We do not study the impact on WMAP data on sterile solutions with  2+2 spectrum,
since they are anyhow incompatible with neutrino data.


\medskip

\paragraph{Conclusion} We collect in table~\ref{tab:global2} the
present status of various global interpretations of the solar, atmospheric and LSND neutrino anomalies.
None of them allows to reconcile all neutrino data in a clean way.
It will be interesting to see if MiniBoone will confirm LSND.



\label{out}
\end{document}